\documentclass[runningheads]{llncs}
\usepackage{graphicx}
\usepackage{subfigure}
\usepackage{url}

\usepackage{floatrow}
\usepackage{caption}
\usepackage{pgfplots}
\usepackage{pgfplotstable}
\usepackage{booktabs}
\pgfplotsset{compat=1.13}

\begin{document}
\title{Faster and Smaller Two-Level Index for Network-based Trajectories
\thanks{Funded in part by European Union's Horizon 2020 research and innovation programme
under the Marie Sk\l odowska-Curie grant agreement 690941, by CONICYT-PFCHA/Mag\'isterNacional/2016 - 22161080 (R.R.), and Millennium Institute for Foundational Research on Data and Fondecyt-Conicyt grant number 1170497.}}

%
%
\author{Rodrigo Rivera \and
Andrea Rodr\'iguez \and
Diego Seco}
\authorrunning{R. Rivera et al.}
%
\institute{Department of Computer Science, University of Concepci\'on, Chile\\
\email{\{rodrivera,andrea,dseco\}@udec.cl}}
\maketitle              
\begin{abstract}
Two-level indexes have been widely used to handle trajectories of moving objects that are constrained to a network. The top-level of these indexes handles the spatial dimension, whereas the bottom level handles the temporal dimension. The latter turns out to be an instance of the \textit{interval-intersection} problem, but it has been tackled by non-specialized spatial indexes. In this work, we propose the use of a compact data structure on the bottom level of these indexes. Our experimental evaluation shows that our approach is both faster and smaller than existing solutions.

\end{abstract}
\section{Introduction}

Spatio-temporal information has gained popularity in decision making systems, such as optimization of transportation systems, urban planning, and so on. The proliferation of different types of sensors to capture or generate this kind of data has made these applications possible but, at the same time, it has also made challenging the storage and processing of spatio-temporal data. The work in this paper focuses on a subcategory of spatio-temporal data, that is, trajectory of moving objects, which can be be reconstructed by the GPS devices of smart-phones or, at a different granularity, by smart transportation cards.

Trajectories can be classified as free-trajectories, in which movement is not constrained, and network-based trajectories, in which movement is constrained to a network and cannot exist outside such network. Hurricanes and animal migrations are examples of the former, whereas public transportation is an example of the latter. Useful queries that can be answered by handling trajectories are: count the number of vessels inside a region during a time period (e.g. fishing closed season) or find the shortest path between two stops of a transportation system during a time period. 

Several spatio-temporal indexes have been proposed to handle both free and network-based trajectories. However, classical solutions to deal with moving-object data are inefficient when facing the data volume collected through new sensor technology and the increasing interest for data analysis. On the other hand, space-efficient data structures have been proved to be successful for handling large volumes of data in many different domains, such as the Web, biological sequences, documents and code repositories, to name some examples.

In this work, we focus on two-level indexes for network-based trajectories and propose a new solution that uses compact data structures on the bottom level. This approach turns out to be smaller and faster than existing solutions.

\section{Background and Related Work}

A data structure for trajectories must provide access methods that allow the processing of spatio-temporal queries. These queries can be classified into coordinate- and trajectory-based queries~\cite{TB}.
Coordinate-based queries include \textit{time-slice} queries that determine the position of  objects at a given time instant, \textit{time-interval} queries that  extend  time-slice queries  to a time range, and queries about \textit{nearby neighbors}. 
As for trajectory-based queries, they include  topological queries, which  involve information regarding the movement of an object, and queries related to navigation, which involve information  derived from the movement, such as speed or direction. There also exist combinations such as ``Where was object $X$ at a given time instant''. 

Various data structures have been proposed to efficiently support queries on trajectories. These structures can be broadly classified into two categories: i) Data structures to support free  movements on a space, such as 3D R-tree (a three-dimensional extension of the R-tree \cite {3DR}), TB-tree \cite {TB} (which preserves the trajectories while allowing typical range queries on an R-tree) and MV3R-tree \cite{MVR3} (which uses a multi-version R-tree, called MVR-tree, along with an auxiliary 3D R-tree). ii) Data structures to support movements on networks, such as FNR-tree~\cite {FNR} (which uses a combination of a 2D R-tree with a forest of 1D R-trees), MON-tree~\cite{MON} (using 2 levels of 2D R-trees) and PARINET~\cite {PARINET} (based on graph partitioning and the use of B$^+$-tree). Among the previous structures, FNR-tree and MON-tree have in common the separation of spatial and temporal dimensions, using a  spatial structure (two-dimensional) and a forest of temporal structures (one-dimensional) to tackle each of these sub-problems separately. 

Like FNR-tree and MON-tree, we focus on these two-level indexes. To solve the spatial problem, that is, the representation of the network in space (two-dimensional plane), aforementioned structures use a 2D R-tree, storing the segments of the network as lines. With the spatial problem solved, time has to be associated with segments in the network. More precisely, it is necessary to look for all the time intervals (times in which some objects pass through a segment) that intersect with a given query interval. This problem is known in the literature as \textit{interval intersection}, an extension of the \textit {interval stabbing problem}~\cite{Schmidt2009a}. Classical structures to solve this problem are Interval trees and Priority trees~\cite{stabbing}.

As for the subproblem in the temporal dimension, FNR-tree makes use of a one-dimensional R-tree for each segment. These 1D R-trees index the objects whose trajectories pass through the segments of the network, storing the instant they enter and leave the segment in the form of a time interval $(t_ {entry}, t_ {exit})$. Since only these intervals are stored, the structure assumes that objects do not stop or change speed or direction in the middle of a segment, they can only do so at nodes. MON-tree eliminates this restriction by replacing the one-dimensional R-trees with two-dimensional R-trees, where they store the relative movement within the segments as rectangles in the 2D R-tree of the form $(p_1, p_2, t_1, t_2)$, with $ (p_1, p_2) $ a range of relative positions and $ (t_1, t_2) $ a temporal interval. 

While some of aforementioned structures support queries efficiently on large datasets, they are incapable of handling the increasing data volume of current applications. This has forced the use of compression techniques for data storage and transmission. Some techniques are to reduce the number of points in a curve~\cite{PTS} or to use features at each point, such as speed and orientation~\cite {PVO}. Both techniques work in free spaces and, when the movement is restricted to networks, it is even possible to get a better compression, like the ones shown in~\cite{COMPRESS,CR2,CiNCT,CR1}.

Previous compression techniques improve storage requirements and transmission time of large datasets. However, the compression can be directly exploited by data structures that can maintain a compact representation of the data while allowing for indexed search capabilities. These structures have been called self-indexes and have been successfully implemented in other domains, such as information retrieval~\cite{self}.

Recently, compact data structures have been also used for the representation of trajectories. GraCT~\cite{GraCT}, for free paths, uses a k$^{2}$-tree~\cite{Brisaboa2009} to store the absolute position of the objects in regular time intervals (snapshots) plus compressed logs  for the representation of the movements between snapshots. ContaCT~\cite{ContaCT} improves GraCT with more efficient logs. Both structures answer spatio-temporal queries where space and time are the main filters, such as, ``finding  trajectories that went through a specific region at a given time instant". On the other hand, CTR~\cite{CTR} supports trajectories restricted to networks by combining compressed suffixes arrays (CSA), to represent the nodes on the network an object passes through, and a balanced Wavelet matrix for the temporal component of the movement. In CTR, trajectories (or trips) are defined as sequences of labels, which represent the nodes of the network. Hence, it solves other types of queries in which the space is represented with such labels, such as ``find the number of trajectories that started at $X$ and ended at $Y$". This is a fundamental difference with our proposal, in which the spatial dimension are coordinates in a two-dimensional space, and not labels. This is also the main difference with CiNCT \cite{CiNCT}, which boosts CTR in terms of memory storage and query time.

Another difference with previous solutions is that our approach uncouples the network from the trajectories. This model known as Network-Matched has been successfully used~\cite{zhiming15,krogh14}, but without using compact data structures in its implementation. Our approach has the advantage that mapping trajectories to a network facilitates the finding of similar trajectories and, in consequence, it allows a better use of space.

\section{Data Structures for Network-based Trajectories} \label{sec:data_structures}

Similarly to the FNR-tree and MON-tree, we propose an index with two levels: spatial (top level) and temporal (bottom level). In a preliminary experimental evaluation, we observed that the spatial level requires negligible space compared with the temporal level. For example, for the Oldenburg network (see Section \ref{sec:experimental_results}), in the baseline structure, the temporal level uses about 89\% of the total memory with 1,000 objects circulating and about 94\% with 2,000 objects. The more data are stored, the more negligible the spatial level becomes (due to the almost-static nature of the transportation network in comparison with the moving objects). Hence, we focus on optimizing the temporal level and process the spatial level with a two-dimensional R-tree, as the FNR-tree and MON-tree do. Recall that the R-tree is a balanced tree in which each leaf stores an entry of the form (id, MBB), where id is a reference to the data (in this case to a temporal index) and MBB is the Minimum Bounding Box that covers the spatial object (a line segment, in this domain). The R-tree does not provide worst-case guarantees as it may be forced to examine the entire tree in $O(n)$ time, even when the output is empty. However, it performs well in practice and is ubiquitous in spatial databases.

Each leaf of the R-tree contains a reference to a temporal index. These indexes solve the \textit{Interval Intersection} problem. Before presenting alternatives to solve this problem, we give an overview of a query algorithm for spatio-temporal range queries, which are the most general coordinate-bases queries. First, a spatial query, a 2D window, is solved on the 2D R-tree, which returns a set of leaves whose segment may intersect the window. As in most spatial indexes, a refinement step is then executed to eliminate false positives, i.e. network segments whose MBBs intersect the window, but they do not actually intersect the window. After this refinement, the interval intersection query is executed in each temporal index referenced by the remaining leaves of the R-tree. Results from all these temporal indexes are then combined using an implementation of a set.

\subsection{Temporal Level: Data structures for the Interval Intersection Problem}

Unlike the FNR-tree and MON-tree, which use variants of an R-tree, we explore the use of specialized data structures for the interval-intersection problem.

\subsubsection{Interval-tree \cite{stabbing}.}
This is a binary tree that is constructed recursively in the following way: i) The median $x_{med}$ of all the interval endpoints is computed. ii) Intervals are classified in three sets, $I_ {med}$, $\mathit{I_{left}}$ and $I_{right}$, which contain intervals stabbed by $x_{med}$, intervals to the left of $x_{med}$ and intervals to the right of $x_{med}$, respectively. iii) $I_ {med}$ is stored in a structure composed of two arrays sorted by left and right endpoints, and associated with the root, whereas $\mathit{I_{left}}$ and $I_{right}$ are recursively processed and assigned as left and right child, respectively.

A search for the intervals that intersect with the query interval $(l_{q}, r_ {q})$ is solved recursively starting from the root. The intervals within the visiting node that intersect the query interval are returned and the search is continued in the left child if $l_ {q}$ is less than $x_ {med}$ and/or in the right child if $r_ {q}$ is greater than $x_ {med}$. This data structure requires linear space and $O(\log n+k)$ query time, where $k$ is the number of reported results.

\subsubsection{Schmidt.} The structure presented in \cite {Schmidt2009a} to solve the Interval Stabbing problem can be extended to solve also the Interval Intersection problem \cite{Brisaboa2013}. It defines the \textit{father} of an interval as the rightmost interval among those that cover it completely. This relation forms a tree where siblings are ordered from left to right, and the root of the tree is a special node that acts as the father of all the intervals that are not covered by any other. In addition, for each possible endpoint of an interval, the structure stores an array called \texttt{start}, with a pointer to the node representing the rightmost-starting interval that intersects such point, and an array \texttt{start2}, storing a pointer to the node representing the rightmost interval starting up to such point (which may not be stabbed by it).

To solve an interval intersection query $q$, the algorithm first reports the rightmost interval that intersects $q$, which is $\max(start[l_q],start_2[r_q])$, if it exists. Then, the algorithm recursively reports the siblings to the left of the node while its right endpoint is greater than or equal to $l_{q}$, also searching among the right children of the reported nodes. This structure requires linear space and optimal $O(1 + k)$ query time. Note, however, that this solution works only for small integer ranges. In order to work with intervals whose endpoints are floats, these endpoints are stored in sorted arrays and two binary searches are used to translate the query to rank space \cite{Brisaboa2013}, which results in a total complexity of $O(\log n+k)$.

\subsubsection{Compact data structure based on Independent Interval Sets (IIS).} \label{subsec:IIS}
A set of intervals $I = \{i_{1},i_{2},...,i_{n}\}$ is called an \textit{Independent Interval Set} if no interval $i_{j} \in I$ is contained in any other interval $i_{k} \in I$.

\begin{figure}[ht]
\centering
\includegraphics[width=0.5\linewidth]{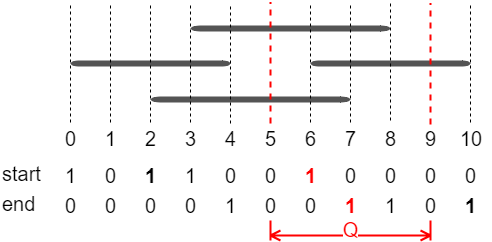}
\caption{An Independent Interval Set (IIS) and its representation with two bitvectors. In red, the last interval stabbed in \texttt{start} and the first one in \texttt{end}.}
\label{fig:bv}
\end{figure}

Report the $k$ intervals of an \textit{IIS} that intersect a query interval $Q = [l_q, r_q] $ can be easily computed if we have the intervals in order. Note that, by definition of \textit{IIS}, the order of the left endpoints of the intervals is the same as that of the right endpoints. If the first and the last interval intersected by the query are located, it is enough to iterate between them to return all the intersected intervals (see Figure \ref{fig:bv}).

In order to locate these two intervals, we could store the left and right coordinates of the intervals in two sorted arrays and use binary search to locate them, which is similar to what we did in previous solution. However, for this domain, we propose a simple solution that facilitates the use of compact data structures. Recall that the endpoints of our intervals are timestamps represented as float numbers. We multiply these timestamps by a scale factor to convert them to integers. For example, if we work with timestamps with up to $6$ decimals it is enough to multiply each one of them by $10^ {6}$ to discretize the space. With this procedure, we obtain integer endpoints in an universe $U$ and, the larger the scale factor, the larger the universe.

After this discretization, we use two bitvectors, one for the left endpoints, \texttt{start}, and another for the right endpoints, \texttt{end}, of each interval in the set (see Figure \ref{fig:bv}). A $1$-bit in these bitvectors indicate that an interval starts (or ends, respectively), at such position. Then, for a query $Q = [l_q, r_q]$ also discretized to this universe, two rank operations on these bitmaps are used to locate the first and last intervals intersected by the query: $rank_1(end, l_q)$ and $rank_1(start, r_q)$, respectively. As we mentioned above, the larger the scale factor, the larger the size of the universe $u$, which is the number of bits in these bitmaps. However, the number of set bits in them is $n$, which is the number of intervals (independently of the scale factor). Hence, we use the Elias-Fano representation \cite{okanohara} for this bitmaps, which takes $2n+n\log\frac{u}{n}$ bits of space. Note that, for a constant $c$ and $u=O(n^c)$, it uses linear space as previous structures. The query time of rank operations on these bitmaps is $O(\log\frac{u}{n})$, thus, this structure can report the $k$ intervals intersecting the query in $O(\log\frac{u}{n}+k)$ time.

Although this solution only works for IIS, a general set of intervals can be decomposed into $m$ independent sets in $O(n\log m)$ time, for example, with Fredman's algorithm \cite{Fredman} to find the optimal number of shuffled upsequences in a permutation (by considering the rightmost endpoints of the intervals as the permuted values). This leads to a solution that requires $O(m\log\frac{u}{n}+k)$ time to report the $k$ solutions. This does not provide worst case guarantees as $m$ can be as large as $n$, however, this adaptive analysis shows that this is an efficient solution for domains in which $m$ is small. The empirical evaluation in next section shows that this is precisely the case in our domain.

\section{Experimental Results} \label{sec:experimental_results}

All the implementations evaluated in this paper were coded in C++11. For the baselines, we use some available implementations: R-tree \cite{code:rtree}, Interval-tree\footnote{This implementation uses sequential search in each node, which is not optimal in theory, but performs well in practice.} \cite{code:intervaltree} and Schmidt \cite{code:schmidt}. We also make use of some succinct data structures from the SDSL library \cite{code:sdsl}. The experiments were run in a computer with an Intel Xeon E3-1220 v5 of 3.00 GHz CPU, 64GB of RAM, and implementations were compiled with g++ 5.4.0 over Ubuntu 16.04 (64 bits).

We first evaluate the performance of all the implementations for interval intersection on synthetic datasets, and then, the best candidates are evaluated in the complete solution for network-based trajectories.

\subsection{Evaluation of Interval Intersection Data Structures}

\begin{figure}[h!]
\noindent
\begin{tikzpicture}
\begin{axis}[
        font = \tiny,
        width = 0.38\textwidth,
        legend pos = north west,
        legend style = {draw = none, fill = none, 
            nodes={scale=0.7}},
        title = a) Query time,
        xlabel = Number of intervals ($\times 10^5$),
        ylabel = Time (s),
        xmin = 1, ymin = 0
    ]
    \addplot coordinates
    {(1,.872395) (2,1.920595) (4,3.948101) (8,8.054299)};
    \addplot coordinates
    {(1,1.195076) (2,2.388729) (4,4.776661) (8,9.548802)};
    \addplot[mark=star] coordinates
    {(1,.845613) (2,1.674556) (4,3.703683) (8,7.491252)};
    \addplot[mark=triangle*] coordinates
    {(1,.078797) (2,.176882) (4,.497121) (8,1.257332)};
    \legend {R-tree, Interval-tree, Schmidt, IIS}
\end{axis}
\end{tikzpicture}
    \hfill
\begin{tikzpicture}
\begin{axis}[
        font = \tiny,
        width = 0.38\textwidth,
        title = b) Space (log scale),
        xlabel = Number of intervals ($\times 10^5$),
        ylabel = Memory usage (MB),
        ymode = log,
        xmin = 1
    ]      
    \addplot coordinates
    {(1,2.852216) (2,5.695840) (4,11.385128) (8,22.895936)};
    \addplot coordinates
    {(1,.802256) (2,1.602448) (4,3.202736) (8,6.402976)};
    \addplot[mark=star] coordinates
    {(1,16.800200) (2,33.600200) (4,67.200200) (8,134.400200)};
    \addplot[mark=triangle*] coordinates
    {(1,.401742) (2,.709528) (4,1.213136) (8,1.955322)};
\end{axis}
\end{tikzpicture}
    \hfill
\begin{tikzpicture}
\begin{axis}[
        font = \tiny,
        width = 0.38\textwidth,
        title = c) Construction time,
        xlabel = Number of intervals ($\times 10^5$),
        ylabel = Time (s),
        xmin = 1, ymin = 0
    ]   
    \addplot coordinates
    {(1,.398296) (2,.825573) (4,1.740655) (8,3.642007)};
    \addplot coordinates
    {(1,.149329) (2,.286914) (4,.579077) (8,1.165272)};
    \addplot[mark=star] coordinates
    {(1,.238883) (2,.695869) (4,1.416728) (8,2.892020)};
    \addplot[mark=triangle*] coordinates
    {(1,.733202) (2,1.797077) (4,1.996310) (8,2.712820)};
\end{axis}
\end{tikzpicture}
\caption{Fixed size intervals.}
\label{gra:definido}
\end{figure}
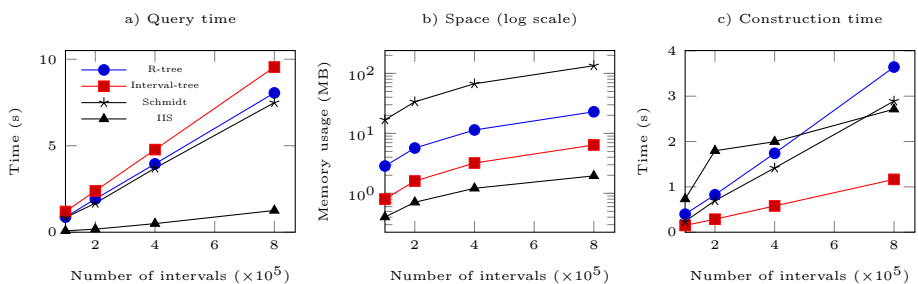

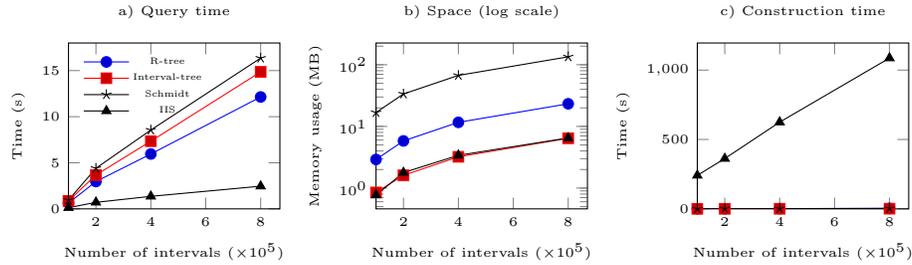
\begin{figure}[h!]
\noindent
\begin{tikzpicture}
\begin{axis}[
        font = \tiny,
        width = 0.36\textwidth,
        legend pos = north west,
        legend style = {draw = none, fill = none, 
            nodes={scale=0.7}},
        title = a) Query time,
        xlabel = Number of intervals ($\times 10^5$),
        ylabel = Time (s),
        xmin = 1, ymin = 0
    ]
    \addplot coordinates
    {(1,.621210) (2,2.963652) (4,5.953411) (8,12.140528)};
    \addplot coordinates
    {(1,.862714) (2,3.672076) (4,7.341250) (8,14.861710)};
    \addplot[mark=star] coordinates
    {(1,.956470) (2,4.403502) (4,8.551824) (8,16.357282)};
    \addplot[mark=triangle*] coordinates
    {(1,.124798) (2,.707447) (4,1.358245) (8,2.458945)};
    \legend {R-tree, Interval-tree, Schmidt, IIS}
    
\end{axis}
\end{tikzpicture}
    \hfill
\begin{tikzpicture}
\begin{axis}[
        font = \tiny,
        width = 0.36\textwidth,
        title = b) Space (log scale),
        xlabel = Number of intervals ($\times 10^5$),
        ylabel = Memory usage (MB),
        ymode = log,
        xmin = 1
    ]       
    \addplot coordinates
    {(1,2.907160) (2,5.819464) (4,11.639992) (8,23.268128)};
    \addplot coordinates
    {(1,.848768) (2,1.609072) (4,3.212912) (8,6.417904)};
    \addplot[mark=star] coordinates
    {(1,16.800200) (2,33.600200) (4,67.200200) (8,134.400200)};
    \addplot[mark=triangle*] coordinates
    {(1,.771064) (2,1.783488) (4,3.396936) (8,6.501980)};

\end{axis}
\end{tikzpicture}
    \hfill
\begin{tikzpicture}
\begin{axis}[
        font = \tiny,
        width = 0.36\textwidth,
        title = c) Construction time,
        xlabel = Number of intervals ($\times 10^5$),
        ylabel = Time (s),
        xmin = 1, ymin = 0
    ]      
    \addplot coordinates
    {(1,.412858) (2,.860263) (4,1.794467) (8,3.773696)};
    \addplot coordinates
    {(1,.166731) (2,.281781) (4,.579322) (8,1.165821)};
    \addplot[mark=star] coordinates
    {(1,.344162) (2,.752000) (4,1.554363) (8,3.142946)};
    \addplot[mark=triangle*] coordinates
    {(1,241.158409) (2,362.269212) (4,624.741017) (8,1086.358164)};
    
\end{axis}
\end{tikzpicture}
\caption{Random size intervals.}
\label{gra:aleatorios}
\end{figure}

\begin{figure}[h!]
\noindent
\begin{tikzpicture}
\begin{axis}[
        font = \tiny,
        width = 0.36\textwidth,
        legend pos = north west,
        legend style = {draw = none, fill = none, 
            nodes={scale=0.7}},
        title = a) Query time,
        xlabel = Number of intervals ($\times 10^5$),
        ylabel = Time (s),
        xmin = 1, ymin = 0
    ]
    \addplot coordinates
        {(1,.589466) 
        (2,1.375613) 
        (4,2.855298) 
        (8,5.776594)};
    \addplot coordinates
        {(1,.858606) 
        (2,1.699038) 
        (4,3.334118) 
        (8,6.548112)};
    \addplot[mark=star] coordinates
        {(1,.887573) 
        (2,1.887482) 
        (4,3.965281) 
        (8,8.098209)};
    \addplot[mark=triangle*] coordinates
        {(1,.055175) 
        (2,.124442) 
        (4,.270287) 
        (8,.550162)};
    \legend {R-tree, Interval-tree, Schmidt, IIS}
    
\end{axis}
\end{tikzpicture}
    \hfill
\begin{tikzpicture}
\begin{axis}[
        font = \tiny,
        width = 0.36\textwidth,
        legend pos = south east,
        legend style = {draw = none, fill = none, 
            nodes={scale=0.7}},
        title = b) Space (log scale),
        xlabel = Number of intervals ($\times 10^5$),
        ylabel = Memory usage (MB),
        ymode = log,
        xmin = 1
    ]      
    \addplot coordinates
        {(1,2.891928) 
        (2,5.795392) 
        (4,11.613880) 
        (8,23.182992)};
    \addplot coordinates
        {(1,.947984) 
        (2,1.827088) 
        (4,3.491888) 
        (8,6.770368)};
    \addplot[mark=star] coordinates
        {(1,16.800200) 
        (2,33.600200) 
        (4,67.200200) 
        (8,134.400200)};
    \addplot[mark=triangle*] coordinates
        {(1,.466970) 
        (2,.933664) 
        (4,1.789404) 
        (8,3.422438)};

\end{axis}
\end{tikzpicture}
    \hfill
\begin{tikzpicture}
\begin{axis}[
        font = \tiny,
        width = 0.36\textwidth,
        legend pos = north west,
        legend style = {draw = none, fill = none, 
            nodes={scale=0.7}},
        title = c) Construction time,
        xlabel = Number of intervals ($\times 10^5$),
        ylabel = Time (s),
        xmin = 1, ymin = 0
    ]      
    \addplot coordinates
        {(1,.277140) 
        (2,.574980) 
        (4,1.778141) 
        (8,3.726397)};
    \addplot coordinates
        {(1,.127714) 
        (2,.255174) 
        (4,.740653) 
        (8,1.494681)};
    \addplot[mark=star] coordinates
        {(1,.232926) 
        (2,.475805) 
        (4,1.420820) 
        (8,2.886170)};
    \addplot[mark=triangle*] coordinates
        {(1,3.069251) 
        (2,3.992185) 
        (4,7.154455) 
        (8,10.154502)};
    
\end{axis}
\end{tikzpicture}
\caption{Intervals from trajectories.}
\label{gra:trayectorias_A}
\end{figure}
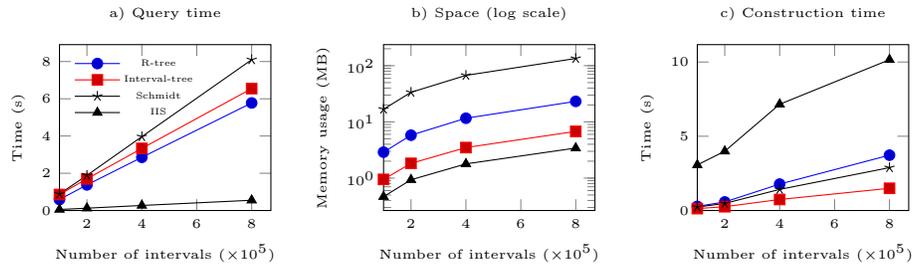

We evaluated the performance in three scenarios with different types of intervals: i) fixed size (Figure \ref{gra:definido}), ii) random size (Figure \ref{gra:aleatorios}), and iii) intervals of trajectories extracted from a trajectories dataset generated with Brinkhoff's generator \cite{brinkoff} over San Francisco's network (Figure \ref{gra:trayectorias_A}). For each of these scenarios, we created a dataset with 800,000 intervals and a queryset with 500 random queries. Reported query time is the total time to solve all the queries.

Figure \ref{gra:definido} shows the performance of the structures using fixed size intervals. The compact data structure shows the best performance among the four structures, with a considerable advantage in both query time and memory usage. In this scenario, intervals do not fully cover each other (except for precision issues), which produces a low number of independent sets in the IIS structure (only 6 for 800,000 intervals). This explains the outstanding performance of IIS.

Figure \ref{gra:aleatorios} shows the performance of the structures for random size intervals. The compact data structure keeps the best results in query time and memory usage (although in a tie with the Interval-tree) while the building time is drastically increased (up to 900 times the building time of the Interval-tree). This is explained by the high number of independent sets (3,273 for 800,000 intervals), which is caused by the frequency with which intervals fully cover each other.

Figure \ref{gra:trayectorias_A} shows the performance of the structures using time intervals extracted from synthetic trajectories obtained with Brinkhoff's generator. The compact data structure shows a performance in between the two previous cases, but more similar to the first one. This shows the sensibility of the structure to the number of independent sets. In this dataset, intervals of trajectories have often similar length, producing a relatively low number of independent sets (29 for 800,000 intervals). Each temporal index is associated with a segment of the network, and moving objects usually traverse a same segment at a similar speed.

\begin{figure}[h!]
\noindent
\begin{tikzpicture}
\begin{axis}[
        font = \tiny,
        width = 0.36\textwidth,
        legend pos = south east,
        legend style = {draw = none, fill = none, 
            nodes={scale=0.85}},
        title = a) Query time,
        xlabel = Scale,
        ylabel = Time (s),
        xmin = 0, ymin = 0
    ]
    \addplot coordinates
    {(0,8.463191) (2,7.034745) (4,6.992891) (6,6.650946) (8,6.935613)};
    \addplot coordinates
    {(0,7.153735) (2,6.930555) (4,7.006115) (6,7.045837) (8,7.002531)};
    \addplot[mark=star] coordinates
    {(0,1.862652) (2,2.212434) (4,7.262291) (6,7.272938) (8,7.277176)};
    \addplot[mark=triangle*] coordinates
    {(0,.294273) (2,.434439) (4,.580004) (6,.561500) (8,.569298)};
    \legend {R-tree, Interval-tree, Schmidt, IIS}
    
\end{axis}
\end{tikzpicture}
    \hfill
\begin{tikzpicture}
\begin{axis}[
        font = \tiny,
        width = 0.36\textwidth,
        title = b) Space (log scale),
        xlabel = Scale,
        ylabel = Memory usage (MB),
        ymode = log,
        xmin = 0
    ]      
     \addplot coordinates
    {(0,35.922112) (2,25.763184) (4,23.205024) (6,23.202168) (8,23.202576)};
    \addplot coordinates
    {(0,6.434848) (2,6.657280) (4,6.770224) (6,6.770416) (8,6.770416)};
    \addplot[mark=star] coordinates
    {(0,134.400200) (2,134.400200) (4,134.400200) (6,134.400200) (8,134.400200)};
    \addplot[mark=triangle*] coordinates
    {(0,.001304) (2,.081246) (4,1.466840) (6,2.835802) (8,4.206066)};

\end{axis}
\end{tikzpicture}
    \hfill
\begin{tikzpicture}

\begin{axis}[
        font = \tiny,
        width = 0.36\textwidth,
        title = c) Construction time,
        xlabel = Scale,
        ylabel = Time (s),
        xmin = 0, ymin = 0, ymax = 6.000000 
    ]      
     \addplot coordinates
    {(0,3.768467) (2,4.045066) (4,4.042143) (6,4.042762) (8,4.083732)};
    \addplot coordinates
    {(0,1.085698) (2,1.459333) (4,1.568968) (6,1.621288) (8,1.633426)};
    \addplot[mark=star] coordinates
    {(0,1.933967) (2,2.473946) (4,3.169172) (6,3.175128) (8,3.219359)};
    \addplot[mark=triangle*] coordinates
    {(0,.564067) (2,.814954) (4,1.022978) (6,2.034129)}; 

\end{axis}
\end{tikzpicture}
\caption{Performance according the scale of the intervals. The last point of IIS in the last graph was omitted, because it is about 40 times larger than the others.}
\label{gra:precision}
\end{figure}
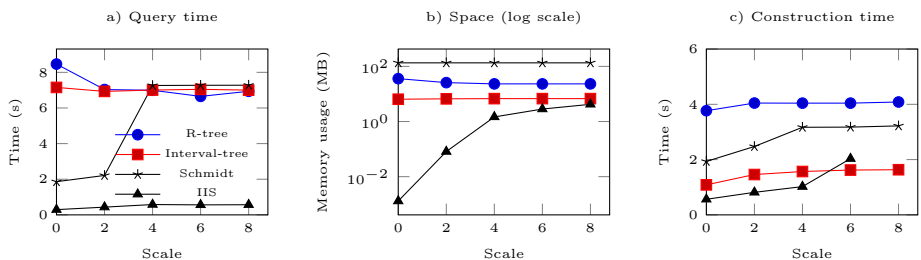

We also evaluated the sensibility of the structures to the scale used to transform original float-number times to integers. In this procedure, each time is multiplied by a scale factor and then truncated. In our datasets, original times use up to 8 digits to the right of the decimal point. Hence, a scale factor of $10^8$ guarantees a lossless transformation, whereas lower scale factors may produce a lossy transformation. In these experiments, we used the same 800,000 intervals of trajectories of the previous evaluation and results are shown in Figure \ref{gra:precision}.

Query time shows an almost constant behavior, except for the increase suffered by Schmidt, which is caused by the high number of duplicates when only 2 or less digits are used for the fractional part. In terms of space and construction time, the compact data structure is more sensible than the other structures, which is caused by the scale process. As we explain in previous section, the larger the scale factor, the larger the size of the bitmaps in this structure. Even so, this structure obtains the best results in both query time and memory usage, also giving the possibility to improve the performance in applications where the user can afford losing some precision. Note, however, that in all the other experiments we consider all the decimals, which is the worst case for our proposal.

\subsection{Overall evaluation}

From the experiments in previous section, we conclude that Schmidt's structure is always outperformed by the others, and thus it is not considered in the implementation of data structures for trajectories. In the following experiments we compare our proposal, based on compact data structures, with two baselines: the original FNR-tree and an ad-hoc baseline in which 1D R-trees are replaced by interval trees. Note that in these experiments we are comparing three two-level indexes, all of them using a 2D R-tree on the top level.

The datasets of trajectories were created using Brinkhoff's generator \cite{brinkoff} over the real road networks of Oldenburg and San Francisco. The former consists of 6,105 nodes and 7,305 edges, whereas the latter consists of 175,343 nodes and 223,343 edges. We created trajectories for 1,000, 2,000, 3,000, 4,000 and 5,000 objects during 100 units of time for both networks.

\subsubsection{Memory usage.}

Figure \ref{gra:mem_total_final} shows the space required by each of the structures. The proposed space-efficient solution (labeled as IIS in the graphs) obtained the best results in all the experiments. In addition, the larger the number of objects moving over the network, the larger the advantage of this structure over the baselines. For small number of moving objects, the total space used by the data structures is dominated by the spatial level, however, as this number increases, the temporal level dominates, and our proposal takes more advantage.

\begin{minipage}[b]{.6\textwidth}
\begin{tikzpicture}
\begin{axis}[
        font = \tiny,
        width = .6\textwidth,
        legend pos = north west,
        legend style = {draw = none, fill = none, 
            nodes={scale=0.85}},
        title = Oldenburg,
        xlabel = Num. of objects ($\times 10^3$),
        ylabel = Memory usage (MB),
        xmin = 0, ymin = 0
    ]
    \addplot coordinates
    {(1.000, 5.409096
    ) (2.000, 10.588808
    ) (3.000, 15.554416
    ) (4.000, 20.211600
    ) (5.000, 24.361808
    )};
    \addplot coordinates
    {(1.000, 4.765792
    ) (2.000, 8.945504
    ) (3.000, 12.935472
    ) (4.000, 16.672192
    ) (5.000, 19.999184
    )};
    \addplot[mark=triangle*] coordinates
    {(1.000, 3.959532
    ) (2.000, 5.935140
    ) (3.000, 7.408376
    ) (4.000, 8.772656
    ) (5.000, 9.870120
    )};
    \legend {FNR-tree, baseline, IIS}
\end{axis}
\end{tikzpicture}
\begin{tikzpicture}
\begin{axis}[
        font = \tiny,
        width = .6\textwidth,
        legend pos = north west,
        legend style = {draw = none, fill = none, 
            nodes={scale=0.85}},
        title = San Francisco,
        xlabel = Num. of objects ($\times 10^3$),
        xmin = 0, ymin = 0
    ]      
    \addplot coordinates 
    {(1.000, 53.416704
    ) (2.000, 84.678944
    ) (3.000, 116.692464
    ) (4.000, 148.810184
    ) (5.000, 180.906464
    )};
    \addplot coordinates 
    {(1.000, 59.548536
    ) (2.000, 84.558328
    ) (3.000, 110.169144
    ) (4.000, 135.863320
    ) (5.000, 161.540344
    )};
    \addplot[mark=triangle*] coordinates
    {(1.000, 39.655332
    ) (2.000, 52.039872
    ) (3.000, 62.986808
    ) (4.000, 74.384700
    ) (5.000, 84.925548
    )};
    \legend {FNR-tree, baseline, IIS}				
\end{axis}
\end{tikzpicture}
\captionof{figure}{Total memory usage.}
\label{gra:mem_total_final}
\end{minipage}
\hfill
\begin{minipage}[b]{.35\textwidth}
\centering
\small
\begin{tabular}{lcc}
\toprule
Structure & Old. & S.F. \\
\midrule
FNR-tree   & 5 & 32                      \\
baseline    & 4 & 26                    \\
IIS     & 1.5 & 11                    \\ 
\bottomrule
\end{tabular}
\captionof{table}{Memory usage per object (KB / object) [5,000 objects and 100 time units].}
\label{tab:mem_obj_final}
\end{minipage}

The approximated memory usage per object is shown in Table \ref{tab:mem_obj_final}, which shows that our approach requires about $70\%$ less memory than the FNR-tree, and about $60\%$ less memory than the baseline, when there are 5,000 objects moving over the networks. The difference between the two datasets is explained by the size of the network, the San Francisco network being much larger. First, part of the space charged to each object is due to the spatial index. However, the size of the network has also an impact on the distribution of objects per edge of the network. As this distribution is very skewed, the larger the network, the larger the number of nodes with few objects, which means an overhead.

\subsubsection{Query Time.}

The time performance of the structures was evaluated for three types of queries, which are the same used in the original evaluation of the FNR-tree \cite{FNR}: i) \textit{Range Queries with Equal Spatial and Temporal Extent}, such as ``find all objects within a given area during a given time interval''; ii) \textit{Range Queries with Larger Temporal Extent}, which query for very large time intervals, including intervals expanding the whole temporal dimension, such as ``find all the objects having ever passed through a given area''; and iii) \textit{Time Slice Queries}, that only consider a time instant, such as ``find all the objects that were in a given area at a given time instant''. For each of these scenarios, we created three query-sets with 500 random queries for each network. 

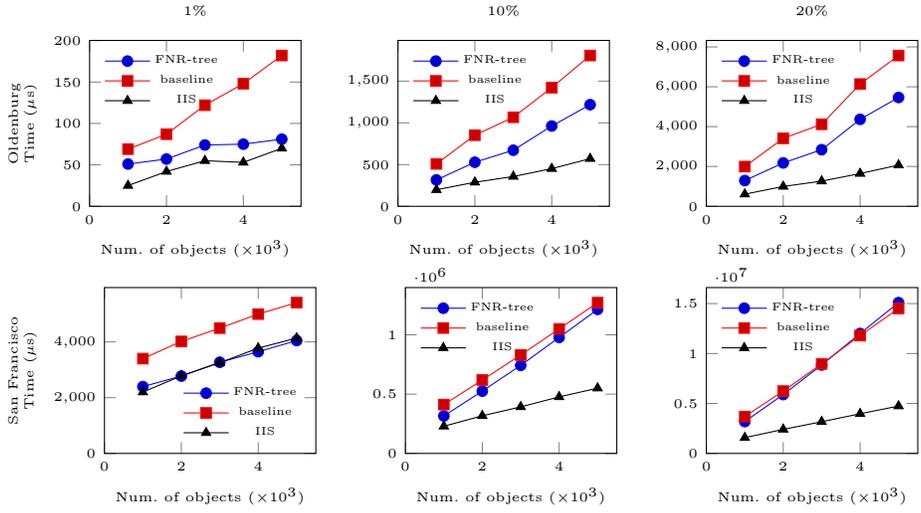
\begin{figure}[h!]%
\noindent
\begin{tikzpicture}
\begin{axis}[
        font = \tiny,
        width = 0.36\textwidth,
        legend pos = north west,
        legend style = {draw = none, fill = none, 
            nodes={scale=0.85}},
        title = 1\%,
        xlabel = Num. of objects ($\times 10^3$),
        style={align=center}, ylabel = Oldenburg\\Time ($\mu$s),
        xmin = 0, ymin = 0
    ]
    \addplot coordinates 
    {(1.000, 51
    ) (2.000, 57
    ) (3.000, 74
    ) (4.000, 75
    ) (5.000, 81
    )};
    \addplot coordinates
    {(1.000, 69
    ) (2.000, 87
    ) (3.000, 122
    ) (4.000, 148
    ) (5.000, 182
    )};
    \addplot[mark=triangle*] coordinates
    {(1.000, 25
    ) (2.000, 42
    ) (3.000, 55
    ) (4.000, 53
    ) (5.000, 70
    )};
    \legend {FNR-tree, baseline, IIS}
\end{axis}
\end{tikzpicture}
    \hfill
\begin{tikzpicture}
\begin{axis}[
        font = \tiny,
        width = 0.36\textwidth,
        legend pos = north west,
        legend style = {draw = none, fill = none, 
            nodes={scale=0.85}},
        title = 10\%,
        xlabel = Num. of objects ($\times 10^3$),
        xmin = 0, ymin = 0
    ]      
    \addplot coordinates 
    {(1.000, 316
    ) (2.000, 528
    ) (3.000, 672
    ) (4.000, 962
    ) (5.000, 1218
    )};
    \addplot coordinates
    {(1.000, 508
    ) (2.000, 852
    ) (3.000, 1067
    ) (4.000, 1422
    ) (5.000, 1807
    )};
    \addplot[mark=triangle*] coordinates
    {(1.000, 198
    ) (2.000, 288
    ) (3.000, 357
    ) (4.000, 451
    ) (5.000, 571
    )};
    \legend {FNR-tree, baseline, IIS}
\end{axis}
\end{tikzpicture}
    \hfill
\begin{tikzpicture}
\begin{axis}[
        font = \tiny,
        width = 0.36\textwidth,
        legend pos = north west,
        legend style = {draw = none, fill = none, 
            nodes={scale=0.85}},
        title = 20\%,
        xlabel = Num. of objects ($\times 10^3$),
        xmin = 0, ymin = 0
    ]      
    \addplot coordinates 
    {(1.000, 1292
    ) (2.000, 2178
    ) (3.000, 2842
    ) (4.000, 4365
    ) (5.000, 5463
    )};
    \addplot coordinates
    {(1.000, 1993
    ) (2.000, 3412
    ) (3.000, 4117
    ) (4.000, 6138
    ) (5.000, 7570
    )};
    \addplot[mark=triangle*] coordinates
    {(1.000, 609
    ) (2.000, 995
    ) (3.000, 1264
    ) (4.000, 1643
    ) (5.000, 2070
    )};
    \legend {FNR-tree, baseline, IIS}
\end{axis}
\end{tikzpicture}
\begin{tikzpicture}
\begin{axis}[
        font = \tiny,
        width = 0.36\textwidth,
        legend pos = south east,
        legend style = {draw = none, fill = none, 
            nodes={scale=0.85}},
        xlabel = Num. of objects ($\times 10^3$),
        style={align=center}, ylabel = San Francisco\\Time ($\mu$s),
        xmin = 0, ymin = 0
    ]
    \addplot coordinates 
    {(1.000, 2397
    ) (2.000, 2776
    ) (3.000, 3271
    ) (4.000, 3649
    ) (5.000, 4050
    )};
    \addplot coordinates
    {(1.000, 3402
    ) (2.000, 4018
    ) (3.000, 4496
    ) (4.000, 5000
    ) (5.000, 5411
    )};
    \addplot[mark=triangle*] coordinates
    {(1.000, 2201
    ) (2.000, 2784
    ) (3.000, 3254
    ) (4.000, 3779
    ) (5.000, 4138
    )};
    \legend {FNR-tree, baseline, IIS}
\end{axis}
\end{tikzpicture}
    \hfill
\begin{tikzpicture}
\begin{axis}[
        font = \tiny,
        width = 0.36\textwidth,
        legend pos = north west,
        legend style = {draw = none, fill = none, 
            nodes={scale=0.85}},
        xlabel = Num. of objects ($\times 10^3$),
        xmin = 0, ymin = 0
    ]      
    \addplot coordinates 
    {(1.000, 315705
    ) (2.000, 525050
    ) (3.000, 742399
    ) (4.000, 977593
    ) (5.000, 1214279
    )};
    \addplot coordinates
    {(1.000, 411415
    ) (2.000, 618875
    ) (3.000, 829621
    ) (4.000, 1049745
    ) (5.000, 1271118
    )};
    \addplot[mark=triangle*] coordinates
    {(1.000, 228910
    ) (2.000, 317207
    ) (3.000, 392291
    ) (4.000, 476516
    ) (5.000, 550087
    )};
    \legend {FNR-tree, baseline, IIS}
\end{axis}
\end{tikzpicture}
    \hfill
\begin{tikzpicture}
\begin{axis}[
        font = \tiny,
        width = 0.36\textwidth,
        legend pos = north west,
        legend style = {draw = none, fill = none, 
            nodes={scale=0.85}},
        xlabel = Num. of objects ($\times 10^3$),
        xmin = 0, ymin = 0
    ]      
    \addplot coordinates 
    {(1.000, 3187541
    ) (2.000, 5891308
    ) (3.000, 8853657
    ) (4.000, 12003029
    ) (5.000, 15092690
    )};
    \addplot coordinates
    {(1.000, 3687656
    ) (2.000, 6241046
    ) (3.000, 8966167
    ) (4.000, 11796439
    ) (5.000, 14507623
    )};
    \addplot[mark=triangle*] coordinates
    {(1.000, 1575509
    ) (2.000, 2405551
    ) (3.000, 3169583
    ) (4.000, 3970094
    ) (5.000, 4739289
    )};
    \legend {FNR-tree, baseline, IIS}
\end{axis}
\end{tikzpicture}
\caption{Range Queries. First row for Oldenburg and second row for San Francisco. Each column contains queries of different size from 1\% to 20\%.}
\label{gra:t_grupo1}
\end{figure}

Figure \ref{gra:t_grupo1} shows the results for the first type of queries. The first row shows results for Oldenburg and the second row for San Francisco. For both datasets, we show the results of random queries of different sizes, 1\%, 10\% and 20\% in each dimension. Similar frameworks will be used to evaluate the other two types of queries. This is the same experimental setting used in \cite{FNR}.

In all the experiments our proposal outperforms both baselines. Just for small queries, 1\% of the dimensions, the FNR-tree shows competitive results with our proposal. This is more evident in the largest network. The justification is the relative importance of the spatial part of the query with respect to the temporal part, which depends on the size of the network. Also important, our proposal shows better scalability on the number of objects moving through the network.

\begin{figure}[h!]%
\noindent
\begin{tikzpicture}
\begin{axis}[
        font = \tiny,
        width = 0.36\textwidth,
        legend pos = north west,
        legend style = {draw = none, fill = none, 
            nodes={scale=0.85}},
        title = 1\% - 10\%,
        xlabel = Num. of objects ($\times 10^3$),
        style={align=center}, ylabel = Oldenburg\\Time ($\mu$s),
        xmin = 0, ymin = 0
    ]
    \addplot coordinates 
    {(1.000, 47
    ) (2.000, 91
    ) (3.000, 103
    ) (4.000, 132
    ) (5.000, 186
    )};
    \addplot coordinates
    {(1.000, 81
    ) (2.000, 130
    ) (3.000, 150
    ) (4.000, 197
    ) (5.000, 246
    )};
    \addplot[mark=triangle*] coordinates
    {(1.000, 9
    ) (2.000, 29
    ) (3.000, 36
    ) (4.000, 41
    ) (5.000, 61
    )};
    \legend {FNR-tree, baseline, IIS}
\end{axis}
\end{tikzpicture}
    \hfill
\begin{tikzpicture}
\begin{axis}[
        font = \tiny,
        width = 0.36\textwidth,
        legend pos = north west,
        legend style = {draw = none, fill = none, 
            nodes={scale=0.85}},
        title = 1\% - 100\%,
        xlabel = Num. of objects ($\times 10^3$),
        xmin = 0, ymin = 0
    ]      
    \addplot coordinates 
    {(1.000, 176
    ) (2.000, 347
    ) (3.000, 413
    ) (4.000, 587
    ) (5.000, 772
    )};
    \addplot coordinates
    {(1.000, 242
    ) (2.000, 414
    ) (3.000, 550
    ) (4.000, 733
    ) (5.000, 886
    )};
    \addplot[mark=triangle*] coordinates
    {(1.000, 37
    ) (2.000, 84
    ) (3.000, 99
    ) (4.000, 144
    ) (5.000, 180
    )};
    \legend {FNR-tree, baseline, IIS}
\end{axis}
\end{tikzpicture}
    \hfill
\begin{tikzpicture}
\begin{axis}[
        font = \tiny,
        width = 0.36\textwidth,
        legend pos = north west,
        legend style = {draw = none, fill = none, 
            nodes={scale=0.85}},
        title = 10\% - 100\%,
        xlabel = Num. of objects ($\times 10^3$),
        xmin = 0, ymin = 0
    ]      
    \addplot coordinates 
    {(1.000, 2112
    ) (2.000, 4060
    ) (3.000, 5872
    ) (4.000, 8851
    ) (5.000, 11272
    )};
    \addplot coordinates
    {(1.000, 2784
    ) (2.000, 5194
    ) (3.000, 6919
    ) (4.000, 9709
    ) (5.000, 12592
    )};
    \addplot[mark=triangle*] coordinates
    {(1.000, 602
    ) (2.000, 1093
    ) (3.000, 1541
    ) (4.000, 1842
    ) (5.000, 2444
    )};
    \legend {FNR-tree, baseline, IIS}
\end{axis}
\end{tikzpicture}
\begin{tikzpicture}
\begin{axis}[
        font = \tiny,
        width = 0.36\textwidth,
        legend pos = north west,
        legend style = {draw = none, fill = none, 
            nodes={scale=0.85}},
        xlabel = Num. of objects ($\times 10^3$),
        style={align=center}, ylabel = San Francisco\\Time ($\mu$s),
        xmin = 0, ymin = 0
    ]
    \addplot coordinates 
    {(1.000, 5015
    ) (2.000, 7708
    ) (3.000, 9832
    ) (4.000, 13304
    ) (5.000, 15886
    )};
    \addplot coordinates
    {(1.000, 6188
    ) (2.000, 8850
    ) (3.000, 10824
    ) (4.000, 13844
    ) (5.000, 16377
    )};
    \addplot[mark=triangle*] coordinates
    {(1.000, 3559
    ) (2.000, 4766
    ) (3.000, 5468
    ) (4.000, 6541
    ) (5.000, 7423
    )};
    \legend {FNR-tree, baseline, IIS}
\end{axis}
\end{tikzpicture}
    \hfill
\begin{tikzpicture}
\begin{axis}[
        font = \tiny,
        width = 0.36\textwidth,
        legend pos = north west,
        legend style = {draw = none, fill = none, 
            nodes={scale=0.85}},
        xlabel = Num. of objects ($\times 10^3$),
        xmin = 0, ymin = 0
    ]      
    \addplot coordinates 
    {(1.000, 4604
    ) (2.000, 6844
    ) (3.000, 9331
    ) (4.000, 12163
    ) (5.000, 14515
    )};
    \addplot coordinates
    {(1.000, 5580
    ) (2.000, 8055
    ) (3.000, 9980
    ) (4.000, 12344
    ) (5.000, 14653
    )};
    \addplot[mark=triangle*] coordinates
    {(1.000, 3225
    ) (2.000, 4320
    ) (3.000, 5162
    ) (4.000, 6103
    ) (5.000, 6832
    )};
    \legend {FNR-tree, baseline, IIS}
\end{axis}
\end{tikzpicture}
    \hfill
\begin{tikzpicture}
\begin{axis}[
        font = \tiny,
        width = 0.36\textwidth,
        legend pos = north west,
        legend style = {draw = none, fill = none, 
            nodes={scale=0.85}},
        xlabel = Num. of objects ($\times 10^3$),
        xmin = 0, ymin = 0
    ]      
    \addplot coordinates 
    {(1.000, 2533286
    ) (2.000, 5171314
    ) (3.000, 7997882
    ) (4.000, 11202646
    ) (5.000, 14333846
    )};
    \addplot coordinates
    {(1.000, 2379036
    ) (2.000, 4611858
    ) (3.000, 7012504
    ) (4.000, 9532896
    ) (5.000, 11990287
    )};
    \addplot[mark=triangle*] coordinates
    {(1.000, 576622
    ) (2.000, 904377
    ) (3.000, 1331753
    ) (4.000, 1852762
    ) (5.000, 2322194
    )};
    \legend {FNR-tree, baseline, IIS}		
\end{axis}
\end{tikzpicture}
\caption{Range Queries with Larger Temporal Extent. First row for Oldenburg and second row for San Francisco. Each column indicates $x$\% - $y$\%, being $x$ the size of each spatial dimension (1\% or 10\%) and $y$ the size of the time intervals (10\% or 100\%).}
\label{gra:t_grupo2}
\end{figure}
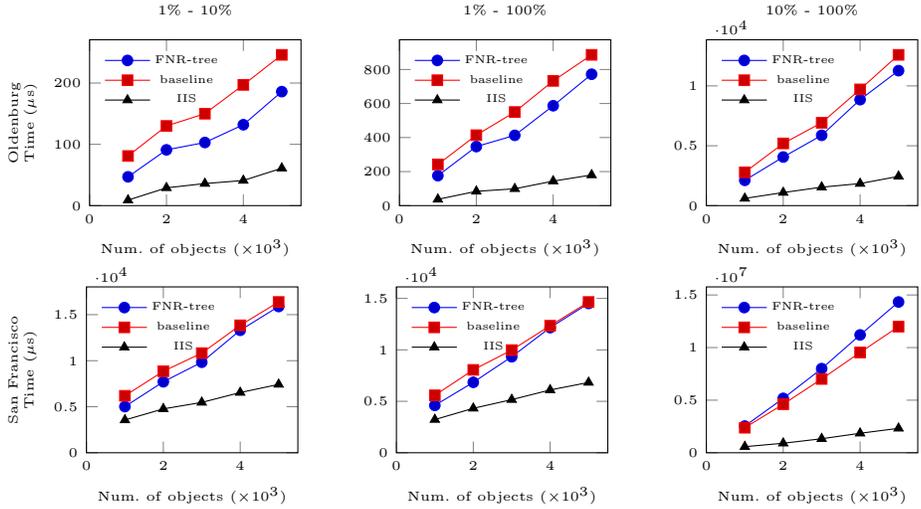

Figure \ref{gra:t_grupo2} shows the results for range queries with larger temporal extent. In these experiments the temporal extent is always larger than the spatial extent, expanding the whole temporal dimension in the second and third column.

\begin{figure}[h!]%
\noindent
\begin{tikzpicture}
\begin{axis}[
        font = \tiny,
        width = 0.36\textwidth,
        legend pos = north west,
        legend style = {draw = none, fill = none, 
            nodes={scale=0.85}},
        title = 1\%,
        xlabel = Num. of objects ($\times 10^3$),
        style={align=center}, ylabel = Oldenburg\\Time ($\mu$s),
        xmin = 0, ymin = 0
    ]
    \addplot coordinates 
    {(1.000, 38
    ) (2.000, 45
    ) (3.000, 66
    ) (4.000, 74
    ) (5.000, 79
    )};
    \addplot coordinates
    {(1.000, 64
    ) (2.000, 80
    ) (3.000, 101
    ) (4.000, 117
    ) (5.000, 152
    )};
    \addplot[mark=triangle*] coordinates
    {(1.000, 6
    ) (2.000, 16
    ) (3.000, 20
    ) (4.000, 20
    ) (5.000, 33
    )};
    \legend {FNR-tree, baseline, IIS}  
\end{axis}
\end{tikzpicture}
    \hfill
\begin{tikzpicture}
\begin{axis}[
        font = \tiny,
        width = 0.36\textwidth,
        legend pos = north west,
        legend style = {draw = none, fill = none, 
            nodes={scale=0.85}},
        title = 10\%,
        xlabel = Num. of objects ($\times 10^3$),
        xmin = 0, ymin = 0
    ]      
    \addplot coordinates 
    {(1.000, 159
    ) (2.000, 275
    ) (3.000, 287
    ) (4.000, 375
    ) (5.000, 464
    )};
    \addplot coordinates
    {(1.000, 365
    ) (2.000, 464
    ) (3.000, 551
    ) (4.000, 695
    ) (5.000, 933
    )};
    \addplot[mark=triangle*] coordinates
    {(1.000, 113
    ) (2.000, 158
    ) (3.000, 238
    ) (4.000, 249
    ) (5.000, 309
    )};
    \legend {FNR-tree, baseline, IIS}
\end{axis}
\end{tikzpicture}
    \hfill
\begin{tikzpicture}
\begin{axis}[
        font = \tiny,
        width = 0.36\textwidth,
        legend pos = north west,
        legend style = {draw = none, fill = none, 
            nodes={scale=0.85}},
        title = 100\%,
        xlabel = Num. of objects ($\times 10^3$),
        xmin = 0, ymin = 0
    ]      
    \addplot coordinates 
    {(1.000, 5464
    ) (2.000, 8013
    ) (3.000, 9549
    ) (4.000, 12843
    ) (5.000, 15448
    )};
    \addplot coordinates
    {(1.000, 14190
    ) (2.000, 19294
    ) (3.000, 23318
    ) (4.000, 30393
    ) (5.000, 36345
    )};
    \addplot[mark=triangle*] coordinates
    {(1.000, 7063
    ) (2.000, 9318
    ) (3.000, 11368
    ) (4.000, 14314
    ) (5.000, 16127
    )};
    \legend {FNR-tree, baseline, IIS}
\end{axis}
\end{tikzpicture}
\begin{tikzpicture}
\begin{axis}[
        font = \tiny,
        width = 0.36\textwidth,
        legend pos = south east,
        legend style = {draw = none, fill = none, 
            nodes={scale=0.85}},
        xlabel = Num. of objects ($\times 10^3$),
        style={align=center}, ylabel = San Francisco\\Time ($\mu$s),
        xmin = 0, ymin = 0
    ]
    \addplot coordinates 
    {(1.000, 2416
    ) (2.000, 2693
    ) (3.000, 2909
    ) (4.000, 3102
    ) (5.000, 3282
    )};
    \addplot coordinates
    {(1.000, 3591
    ) (2.000, 3903
    ) (3.000, 4276
    ) (4.000, 4588
    ) (5.000, 4947
    )};
    \addplot[mark=triangle*] coordinates
    {(1.000, 2525
    ) (2.000, 2871
    ) (3.000, 3307
    ) (4.000, 3752
    ) (5.000, 4033
    )};
    \legend {FNR-tree, baseline, IIS}
\end{axis}
\end{tikzpicture}
    \hfill
\begin{tikzpicture}
\begin{axis}[
        font = \tiny,
        width = 0.36\textwidth,
        legend pos = south east,
        legend style = {draw = none, fill = none, 
            nodes={scale=0.85}},
        xlabel = Num. of objects ($\times 10^3$),
        xmin = 0, ymin = 0
    ]      
    \addplot coordinates 
    {(1.000, 175877
    ) (2.000, 199552
    ) (3.000, 219614
    ) (4.000, 241779
    ) (5.000, 268315
    )};
    \addplot coordinates
    {(1.000, 267781
    ) (2.000, 302803
    ) (3.000, 336428
    ) (4.000, 366753
    ) (5.000, 394889
    )};
    \addplot[mark=triangle*] coordinates
    {(1.000, 211885
    ) (2.000, 274686
    ) (3.000, 325003
    ) (4.000, 374214
    ) (5.000, 423735
    )};
    \legend {FNR-tree, baseline, IIS}
\end{axis}
\end{tikzpicture}
    \hfill
\begin{tikzpicture}
\begin{axis}[
        font = \tiny,
        width = 0.36\textwidth,
        legend pos = south east,
        legend style = {draw = none, fill = none, 
            nodes={scale=0.85}},
        xlabel = Num. of objects ($\times 10^3$),
        xmin = 0, ymin = 0
    ]      
    \addplot coordinates 
    {(1.000, 16567991
    ) (2.000, 18986481
    ) (3.000, 21125620
    ) (4.000, 23193146
    ) (5.000, 25265758
    )};
    \addplot coordinates
    {(1.000, 24165420
    ) (2.000, 27752880
    ) (3.000, 30646261
    ) (4.000, 33545641
    ) (5.000, 36150892
    )};
    \addplot[mark=triangle*] coordinates
    {(1.000, 19714842
    ) (2.000, 25803465
    ) (3.000, 30106132
    ) (4.000, 34554762
    ) (5.000, 38710719
    )};
    \legend {FNR-tree, baseline, IIS}
\end{axis}
\end{tikzpicture}
\caption{Time Slice Queries. First row for Oldenburg and second row for San Francisco. Each columns contains queries of different spatial extent (1\% to 100\%).}
\label{gra:t_grupo3}
\end{figure}
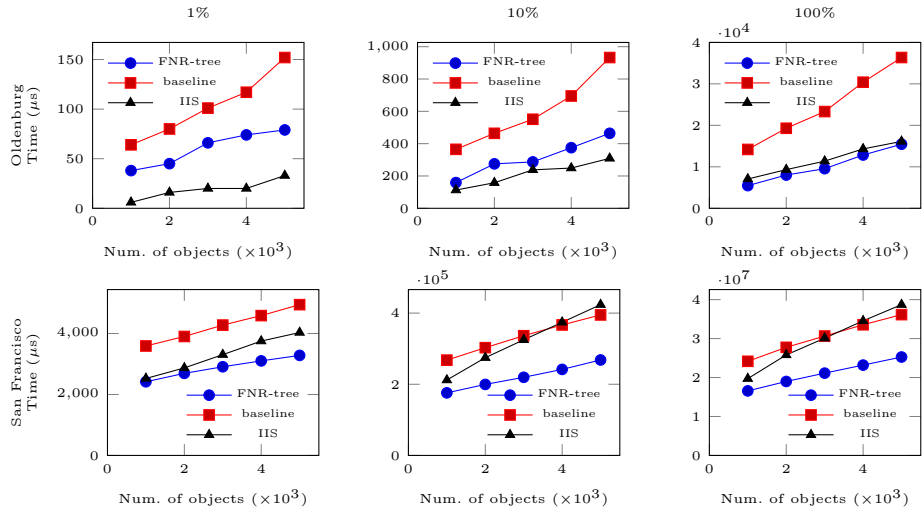

Results in this scenario are similar to the previous one, but the advantage of our proposal is even more obvious. Recall that our structure performs two rank operations in each independent set and then it just iterates over the results, which is very efficient. Finally, Figure \ref{gra:t_grupo3} shows the results for time slice queries.

The analysis of these experiments is quite different from the previous ones, as the FNR-tree usually outperforms all the other approaches. There are two main reasons for this. First, large spatial queries lead querying many temporal indexes (all of them for the experiments in the last column). Second, most of these queries to temporal indexes produce empty results or very few results, which is expensive in our proposal. Each of these queries needs to perform the two rank operations in each independent set just to detect that there are no results to iterate through. Hence, this scenario represents the worst case for our proposal.

\section{Conclusions}

We have proposed a new data structure for trajectories of moving objects, which movement is constrained to a network. Our proposal is inspired by two-level indexes, such as the FNR-tree and MON-tree and, indeed, we use the same two-dimensional R-tree for the spatial dimension. Hence, the difference from previous solutions is in the temporal dimension. This is justified by our experimental evaluation showing that the spatial dimension requires negligible space compared with the temporal dimension. For this dimension, we propose a structure based on a decomposition on independent sets of intervals and the use of succinct data structures. Our experimental evaluation shows that the resulting structure is smaller than previous solutions, and also faster for a broad set of queries.

The interval intersection problem can be reduced to 2-sided range reporting~\cite{Brisaboa2013}, a problem for which efficient data structures have been successfully applied in LZ-indexes~\cite{BelazzouguiCGPR15,BelazzouguiCGPR17}. As these structures are not adaptive to the number of independent interval sets, a combination of both approaches would be interesting as future work. Second, to handle larger datasets, it is necessary to improve construction time. Note, however, that we used larger datasets than those used in the evaluation of the FNR-tree. Third, some parts of the structure could be further optimized. We have observed that the distribution of the moving objects through the network is very skewed, which produces few temporal indexes storing many intervals and many indexes storing very few intervals. Hence, in order to use this index in practice, it is necessary to determine a threshold under which the intervals are just stored in an array and sequentially searched. Finally, bitmaps supporting append operations should be used to support dynamism. 

%
%
\bibliographystyle{splncs04}
\bibliography{mybibliography}

\end{document}